\documentclass{elsart}
\usepackage{graphicx}
\usepackage{epsfig}
\usepackage{amssymb}
\usepackage{color}

\begin{document}
\begin{frontmatter}

\title{Anisotropic response of the moving vortex lattice in superconducting Mo$_{(1-x)}$Ge$_{x}$ amorphous films}

\author{M. I. Dolz$^{a,\ref{x1}}$, D. E. Shal\'om$^{a,\ref{x2}}$, H. Pastoriza$^a$, D. O. L\'opez$^b$}
\address{$^a$ Centro At{\'{o}}mico Bariloche, CONICET, San Carlos de Bariloche, R8402AGP R\'{\i}o Negro, Argentina. \\
$^b$ Center for Nanoscale Materials, Argonne National Laboratory, 9700 South Cass Av., Argonne, IL, 60439, USA.}

\footnote{\label{x1} Present address: Departamento de F\'{\i}sica, Instituto de F\'{\i}sica Aplicada, Universidad Nacional de
San Luis, CONICET, Ej\'ercito de Los Andes 950, D5700BWS San Luis, Argentina.}
\footnote{\label{x2} Present address: Departamento de F\'{\i}sica,Facultad de Ciencias Exactas y Naturales, Universidad Nacional de Buenos Aires, Ciudad Aut\'onoma de Buenos Aires, Argentina.}

\begin{abstract}
We have performed magnetic susceptibility measurements in
Mo$_{(1-x)}$Ge$_x$ amorphous thin films biased with an electrical
current using anisotropic coils. We tested the symmetry of the
vortex response changing the relative orientation between the bias
current and the susceptibility coils. We found a region in the DC
current -- temperature phase diagram where the dynamical vortex
structures  behave anisotropically. In this region the shielding
capability of the superconducting currents measured by the
susceptibility coils is less effective along the direction of
vortex motion compared to the transverse direction. This
anisotropic response is found in the same region where the peak
effect in the critical current is developed. On rising temperature
the isotropic behavior is recovered.
\end{abstract}


\end{frontmatter}

\section{Introduction}

The study of the dynamic behavior of elastic media in the presence
of different static potentials has been center of attention in the
recent years. Superconducting vortices are almost ideal systems
for the study of the phenomenology of these systems because
relevant properties such as density, interactions and external
force can be easily controlled.

Much work has been done on the problem of moving vortex systems.
Depending on the velocity and type of static potential it is
expected to arise different dynamical phases. In the elastic
regime, Giarmarchi and Le Doussal\cite{giamarchi97} have predicted
the existence of a moving-Bragg-glass phase for high driving
currents and a weak disordered static potential. This is a phase
free of dislocations, with a power law decay of the positional
correlations which are anisotropic with respect to flow direction.
On another hand, Balents, Marchetti and
Radzihovsky\cite{balents97} have argued for the existence of a
smectic phase. In this phase, vortices move along well defined
static channels, holding correlation perpendicular to the
channels, but uncorrelated along them. On increasing disorder
plastic motion is expected close to the depinning force showing a
mixture of moving and fixed vortices\cite{gronbechjensen96}. In
the later case the vortex motion is also through meandered static
flow patterns. A common expected feature of all these dynamical
states is the anisotropic response of the moving structure to
perturbations parallel or perpendicular to the vortex motion.

A number of experimental works have been done investigating these
phenomena. Transport measurements report changes in the dynamical
properties of the vortex matter close to the peak effect
\cite{bhattacharya93,henderson96,hellerqvist96,hellerqvist97} that
were attributed to a crossover from plastic to ordered motion.
More recently Kokubo and coworkers\cite{kokubo07} have proven this
dynamical ordering from mode locking experiments. Snapshots of
moving vortex structures have been obtained in decoration
experiments\cite{pardo98,troyanovski99} that have shown evidence
for these smectic and moving glass phases of vortices. However one
of the most striking properties of these quasi-ordered driven
phases is the existence of barriers to a small force transverse to
the direction of motion. These barriers would originate an
effective transverse critical current\cite{fangohr01a} or a change
in the Hall noise spectrum\cite{kolton99}. Although the numerical
results present a clear evidence of these effects the experimental
confirmation is evasive. Recently Lefebvre, Hilke and
Altounian\cite{lefebvre08} have shown transport measurements in
superconducting metallic amorphous tapes that reveal the existence
of a transverse critical current although these results show big
quantitative discrepancies with numerical simulations.

Mutual inductance techniques have proven to be a valuable tool for
measuring superconducting characteristics of two dimensional
systems\cite{hebard80,fiori88}. In previous works it was shown
that a variation of this technique based in special shaped coils
is capable of detecting the anisotropic character of dynamical
phases\cite{shalom03,shalom04}. In this work we extend these
susceptibility measurements to Mo$_{1-x}$Ge$_x$ amorphous films
that reveal the anisotropic characteristic of dynamical vortex
phases. This technique allows the measurement of susceptibility
and transport at the same time obtaining susceptibility
information with applied current to the sample.

\section{Experimental Details}

Mutual inductance measurements were performed using planar
serpentine coils of geometry similar to those described in a
previous paper\cite{shalom04}. It was showed that the
implementation of rectangular coils with high aspect ratio allow
the direct measurement of the anisotropy. Shielding currents are
induced in the sample by the primary coils and therefore they are
predominantly parallel to the long side of the coil, as it is
shown in Figure~1d. As a consequence the voltage generated in the
reception coil is originated by the sample currents flowing in
that direction. For the experiments reported in this work the
coils were batch fabricated on Silicon substrates and later the
superconducting film was deposited on top of them. Coils with
different geometries have been build using 0.5 micron thick
Aluminum as conducting material and SiO$_2$ as insulator. The
fabrication process started from a 8" Silicon wafer covered with
650 nm of Low Stress silicon nitrate. Aluminum wiring of the
primary coil was defined by optical lithography and lift-off.
Deposition of SiO$_2$ by Chemical Vapor Deposition (CVD) with a
nominal thickness of $2\,\mu$m was used as a first isolation
layer. A second lithography step, with the same mask used
previously but rotated 180 degrees defined the secondary coil. A
final SiO$_2$ $2\,\mu$m layer was deposited to provide a second
isolation layer between the coils and the superconducting film. A
planarization with a Chemical Mechanical Polishing (CPM) was
performed afterward. Finally vias in the SiO$_2$ were defined by
lithography and HF etching to allow direct bonding  of Au wires to
each coil. In Figure~1a we show a photograph of one of the
microfabricated coils. All the results presented in this work
correspond to coils with Al wires of $30\,\mu$m in width separated
by $100\,\mu$m and with 30 turns.

On top of the coils the superconducting Mo$_{1-x}$Ge$_x$ 500nm
amorphous films were deposited by RF -- sputtering at room
temperature by water cooling of the sample holder. A cross shaped
geometry was defined in the superconducting film by lithography
and lift-off which allow the DC current injection in the film
parallel and perpendicular to the coil direction. In Figure~1b we
show photograph of the sample and a scheme of the electrical
contacts. Figure~1c presents a sketch of the experimental
configuration. The critical temperature of the films determined as
the onset of resistance at zero magnetic field was $7.98K$.

\begin{figure}
\includegraphics[width=\textwidth]{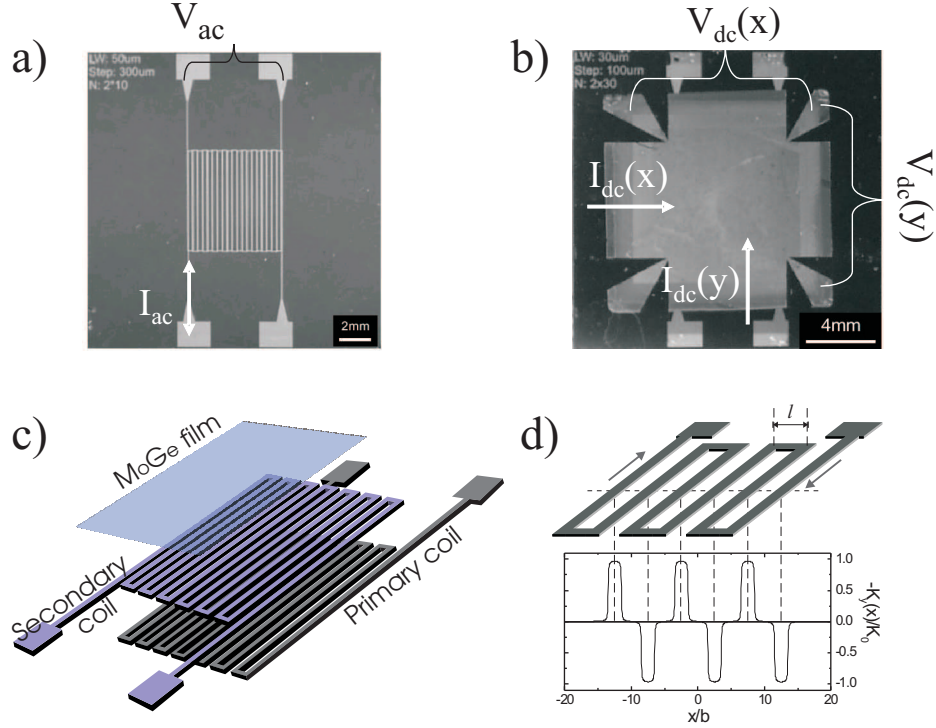}
\caption{ a. Photograph of the microfabricated Al coils. b.
Photograph of the superconducting amorphous film on top of the
serpentine coils. Overlapped are indicated the electrical
contacts. c. Schematics of the measurement configuration. d.
Profile of the current density induced by the primary coil in the
sample.}
\end{figure}

The field and temperature dependence of the susceptibility were
acquired for different applied currents to the sample for both
directions: current parallel ($y$ direction) and perpendicular
($x$ direction) to the long strips of the coils. The voltage drop
in the film was simultaneously measured using a nanovoltmeter that
allows the detection of the vortex motion induced by the external
current. The protocol utilized was the following: for a fixed
external applied field the temperature of the sample was varied
and regulated at fixed setpoints. For each of these temperature
steps the current in the sample was swept in magnitude and
orientation while the susceptibility and voltage drop in the
sample were acquired. All measurements presented here were
performed in the linear regime of the susceptibility and for a
driving frequency and amplitude of $40$ kHz and $0.8$ mA,
respectively.

\section{Results and discussion}

The principal result of this paper is represented in Figure~2
where the susceptibility of the sample is plotted as a function of
temperature, on measuring it with an electrical current through
the sample for the two current orientations and an applied
magnetic field perpendicular to the film. In the following we
define $\chi_\parallel$ as the susceptibility data taken where the
bias current is parallel to the long side of the coils
(\textbf{$I_y$}, $y$ direction), and $\chi_\perp$ where the bias
current on the sample is perpendicular to them (\textbf{$I_x$},
$x$ direction).The data indicate that once the vortex motion is
induced by the external current the response obtained for currents
parallel and perpendicular to the coils are notably different. At
currents lower than 0.1 mA the magnetic response is the same to
both currents orientation. Thus, this magnetic response can be
taken as the reference response, $|\chi'_0|$. At currents greater
than 0.1 mA  the shielding capability, $|\chi'|$, is lower in the
case where the external current is parallel to the long side of
the coils respect to the perpendicular configuration. A lower
value of the $|\chi'|$ data is an indication of an increase in the
vortex mobility. Taking into account that the shielding currents
induced in the sample basically mimic the geometry of the primary
coil \cite{shalom04}, the parallel configuration corresponds to
the case that these shielding and external currents are collinear.
This means that when the bias current is parallel to the long side
of the coils the \emph{dc} force and the alternate testing force
generated by the currents induced in the sample by the primary
coil are almost collinear. As a consequence, the results of these
experiments are evidencing that the vortex mobility is greater
along the vortex motion direction than perpendicular to it. An
isotropic response is recovered at high temperatures where the
data for both current orientations merge together and with that
taken at zero current on the sample. The observed anisotropy
occurs in a region where the susceptibility of reference
$|\chi'_0|$ indicates an increase in the shielding capability. In
Figure 2 we can see this increment from T$=$7.73K, which is an
evidence of the peak effect\cite{chipeakeffect} where the vortex
ensemble softens to accommodate to the weak pinning potential
increasing the critical current.

\begin{figure}
\includegraphics[width=\textwidth]{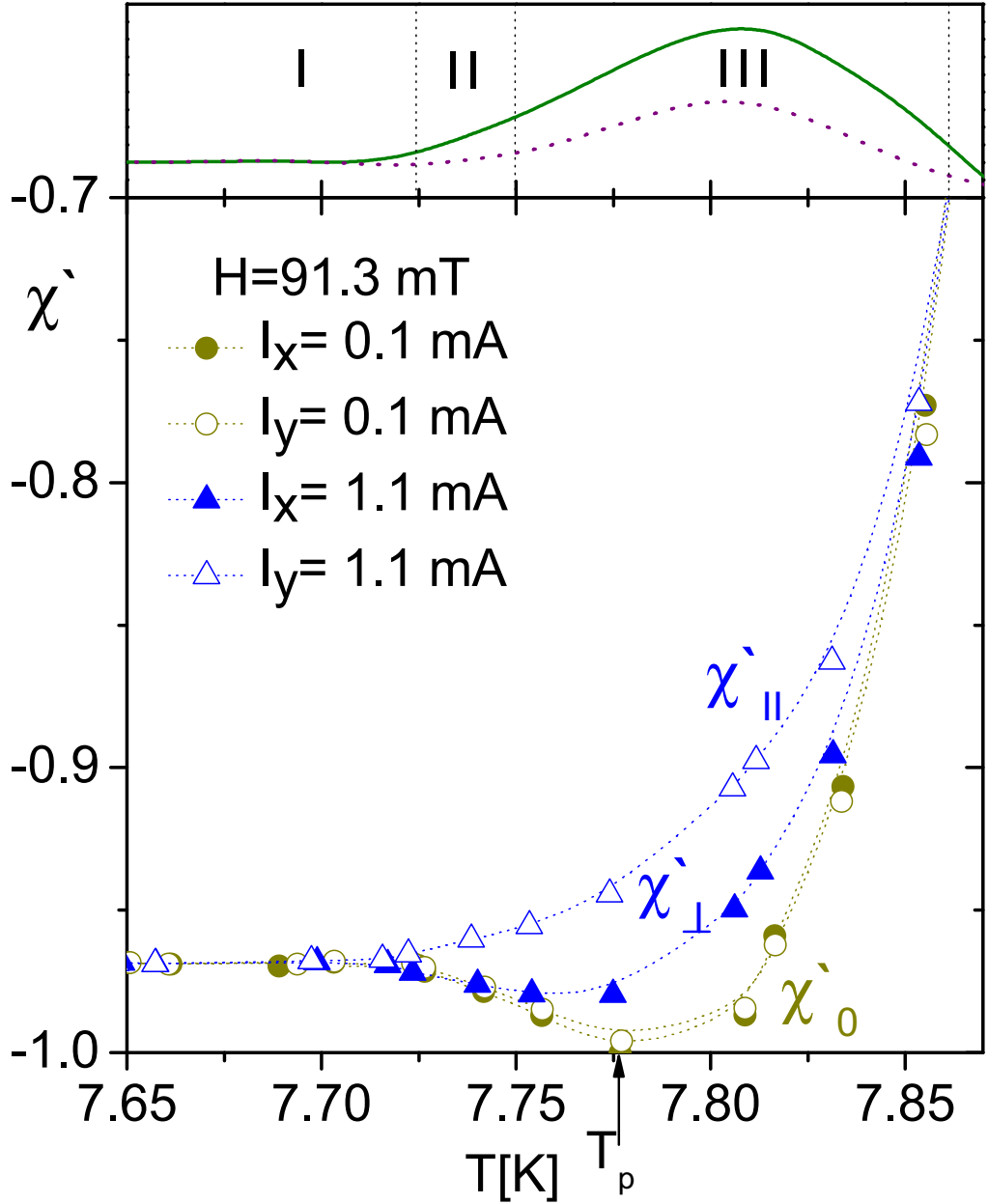}
\caption{Bottom panel: Real part of the susceptibility taken with
two bias currents in the sample of 0.1 mA and 1.1 mA for one
applied magnetic field perpendicular to the film. Open symbols
corresponds to the data taken when the bias current on the sample
is parallel to the long side of the coils. Closed symbols: bias
current perpendicular to the long side of the coils. Dotted lines
are splines to the data. Top panel: Subtraction of the
$\chi_\parallel$ and $\chi_\perp$ fits to the reference $\chi_0$
fit (green and purple line respectively) showing the first three
phases found.}
\end{figure}

In Figure 3 we plot the susceptibility data for fixed temperature
and magnetic field as a function of applied external current in
the sample for both current orientations. The temperature, $T_p$
was chosen as that where the susceptibility of reference
($\chi'_0$) is minimum (peak of the peak effect), see Figure 2. It
is clear that the anisotropic response persists in the whole range
of currents investigated. At this temperature and field the
longitudinal response (or parallel configuration) reaches a steady
state for current values around 3.5 mA. We weren't able to detect
this saturation in the perpendicular configuration up to 5 mA, the
maximum current up to which we were able to measure without
overheating the sample. This effect (difference between parallel
and perpendicular directions) could be in principle explained by
an intrinsic anisotropy of the sample. However this can be
discarded after observing that I -- V characteristics coincide for
both current directions.

\begin{figure}
\includegraphics[width=\textwidth]{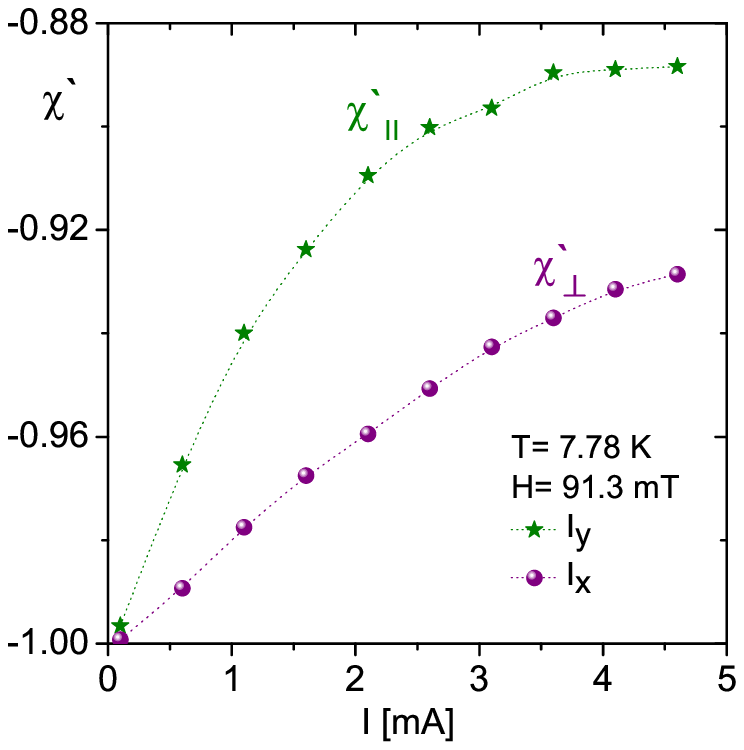}
\caption{Real part of the susceptibility as a function of the bias
current for both current orientations at a fixed value of magnetic
field. The temperature of these measurements was at the minimum of
the susceptibility at zero current, $T_p$, indicated with an arrow
in Figure~2.}
\end{figure}

More insight of the anisotropic behavior can be extracted from the
analysis of the data through a subtraction of the susceptibility
at a given current to a reference measurement at zero current for
both current orientations, see top panel of Figure~2. Based on
this figure (and equivalent for other biasing currents), we were
able to construct the I -- T phase diagram shown in Figure~4. At
low temperatures the susceptibility for both current directions
coincides with the measurement at zero applied current. This
region I corresponds to the response of a fully pinned vortex
lattice. At certain value of the external current we observe that
the susceptibility measured when the current direction is along
the coils long direction starts to deviate from the zero current
behavior but not for the measurement in the perpendicular
configuration. This result indicates that the sample is in a
regime (region II) that the vortex ensemble started to move and is
locked to move in the perpendicular direction proving the
existence of a finite transverse critical current. Rising
temperature the perpendicular susceptibility becomes different
from the zero current measurement but with greater shielding
capability than in the parallel measurement (region III). These
results imply that this is a region of the phase diagram where the
dynamical vortex structure has an anisotropic mobility. At even
higher temperatures and currents an isotropic response is
recovered where all susceptibilities coincide  (region IV) before
reaching the normal state region (V). Region IV could be assigned
as a moving vortex liquid phase where pinning becomes negligible
on the vortex dynamics.

\begin{figure}
\includegraphics[width=\textwidth]{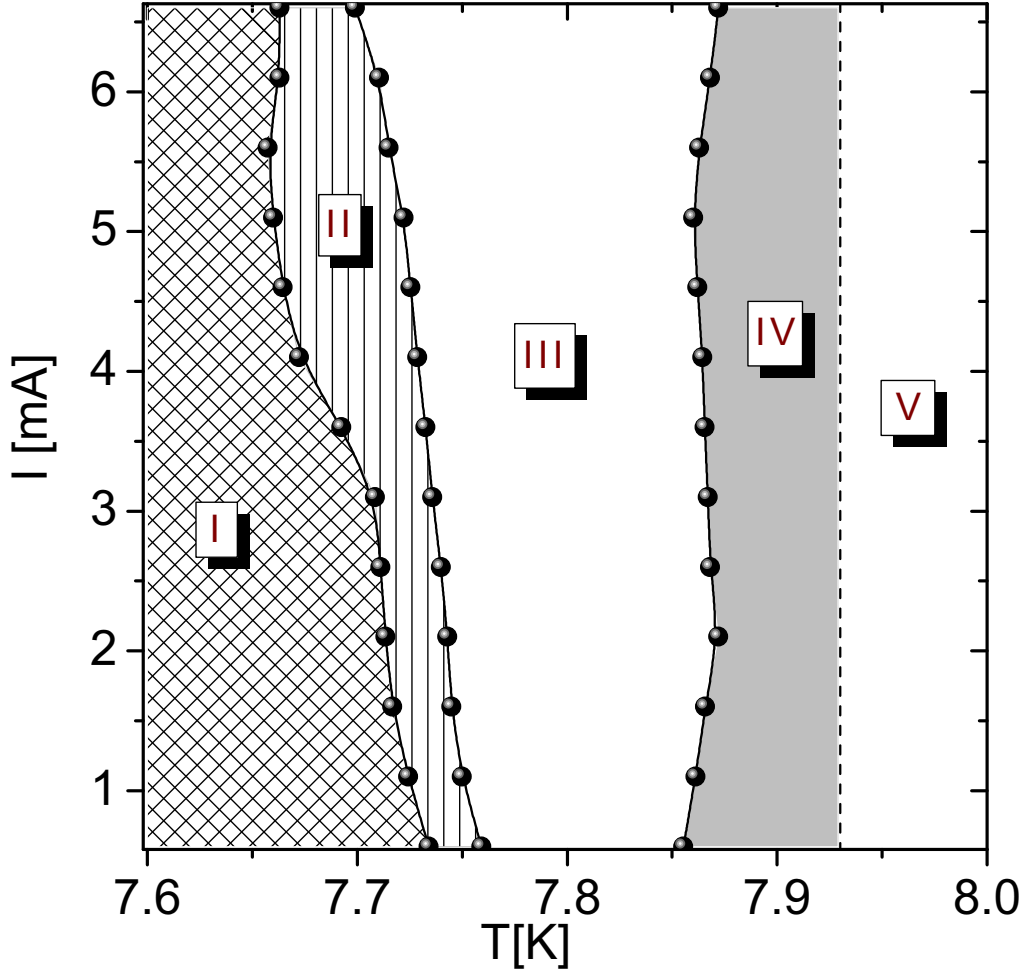}
\caption{Current -- Temperature dynamical phase diagram obtained
from the electrically biased susceptibility measurements. Region
I: Fully pinned state, $\chi_\parallel=\chi_\perp=\chi_0$, $V_{dc}
=0$. Region II: Transversally pinned state, $\chi_\parallel \ne
\chi_\perp=\chi_0$. Region III: Anisotropic dynamical state,
$\chi_\parallel \ne \chi_\perp \ne \chi_0$. Region IV: Isotropic
moving phase: $\chi_\parallel=\chi_\perp=\chi_0$, $V_{dc}
\ne 0$. Region V: Normal state. The phase diagram is relevant only
for currents above $I_{dc}=0.5$ mA, the minimum applied current for
which anisotropy is detected.}
\end{figure}

\section{Conclusions}
Our results clearly indicate that the response of the moving
vortex lattice in $\alpha$-MoGe have an anisotropic characteristic
in the region close to the depinning transition. Our results show
the existence of two different regimes of anisotropic behavior. At
lower temperatures there is a finite response in the $y$ direction
and the response in the perpendicular direction is current
independent. This is a region where there is a transverse critical
current. Rising temperature and/or the magnitude of the current,
the perpendicular component {\em becomes depinned} but the
response is still anisotropic. This anisotropic behavior coincides
in the phase diagram where the peak effect in the critical current
is observed.

\section*{acknowledgments} We are very grateful to M. Hesselberth and P. H.
Kes for helping us with the deposition of the MoGe films on the
planar coils. This project was financially supported by CNEA and
CONICET. M.I.D., D.E.S and H.P. researchers of CONICET.



\end{document}